\documentclass[aps,preprint,prd, showpacs, showkeys]{revtex4-1}

\usepackage{dcolumn}
\usepackage{epsfig,amssymb,multirow,amsmath, bm,xcolor,soul,slashed,}

\begin{document}


\title{Grand canonical potential of a magnetized neutron gas} 



\author{J.P.W. Diener}
	\email{jpwd@sun.ac.za}
 \affiliation{National Institute for Theoretical Physics (NITheP), Stellenbosch 7600, South Africa}
 \affiliation{Institute of Theoretical Physics, University of Stellenbosch, Stellenbosch 7600, South Africa}
\author{F.G. Scholtz}%
 \email{fgs@sun.ac.za}
 \affiliation{National Institute for Theoretical Physics (NITheP), Stellenbosch 7600, South Africa}
 \affiliation{Institute of Theoretical Physics, University of Stellenbosch, Stellenbosch 7600, South Africa}


\date{\today}

\begin{abstract}
We compute the effective action for stationary and spatially constant magnetic fields, when coupled anomalously to charge neutral fermions, by integrating out the fermions.  From this the grand canonical partition function and potential of the fermions and fields are computed.  This also takes care of magnetic field dependent vacuum corrections to the grand canonical potential.  Possible applications to neutron stars are indicated.

\end{abstract}

\pacs{21.65.Cd 21.65.Mn 11.10.Gh }
\keywords{zero-point energy, effective action, magnetized neutron matter}
\maketitle 
\section{Introduction}
The study of magnetized nuclear matter systems finds application in physical systems at various length scales and temperatures: It has been proposed that magnetic fields of the order of $10^{19}$ G can be induced in non-central heavy-ion collisions \cite{sb}. Magnetic fields of similar strength can also be found in the interior of highly magnetized neutron stars, called magnetars \cite{KandK, FandR}.\\
\\
Neutron stars are extremely compact objects with densities similar to, and greater than, that of nuclear matter. These are stable systems and assumed to be in thermal equilibrium at zero temperature, since the thermal energy is much less than the Fermi energy of the particles in the star's interior. Matter with similar densities is also formed in relativistic heavy-ion collisions. Although this type of matter is stable for brief periods, due to the much higher temperatures, it is also assumed to be in thermal equilibrium \cite{TDRHIC}.\\
\\
Under the assumption of thermal equilibrium, the appropriate description of these systems is a statistical description based on the grand canonical ensemble with the grand canonical partition function and closely related potential as the central quantities.  Quite generally, therefore, the computation of the grand canonical potential for nuclear matter at finite temperature and density in the presence of strong magnetic fields is highly relevant. The most convenient formalism for this is finite temperature field theory as described in many textbooks.  Here we follow the formalism as presented in \cite{kapgale}. \\
\\
Due to its high relevance there were a number of recent studies of this quantity using a variety of techniques and focussing on a variety of contributions.  In \cite{cangemi} finite temperature effects for baryons in strong magnetic fields were studied by computing the effective action for the electromagnetic field.  More recently \cite{ferrer} explored the thermodynamical properties of strongly magnetised fermions with emphasis on the issue of pressure anisotropy induced by the magnetic field.  In both these works the anomalous coupling and possible vacuum contributions to the grand canonical potential were not taken into account i.e., the ``no sea'' approximation was made.  It is important to emphasise that the anomalous coupling referred to here is not just the radiative corrections for charged elementary particles, which one expects to have a minor effect \cite{ferrer1}, but it also arises in effective theories that treat structured baryons with an associated magnetic dipole moment as point particles.  To compensate for this it is required to include an anomalous coupling to the dipole moment of such particles in their description.  This is, of course, the case for charged and charge neutral baryons, but it should be clear that it is of particular importance for charge neutral ones, such as neutrons, since it is their only interaction with the electromagnetic field.   The role of this coupling for magnetised fermions was first explored in \cite{Brod00} and has since been used extensively to investigate magnetised systems. Indeed, in a recent study  \cite{strick} the role of an anomalous coupling in determining the bulk properties of a gas of charged and charge neutral fermions was explored quite extensively, again emphasising the role of pressure anisotropy.  However, this study also did not include any possible vacuum contributions to the grand canonical potential.\\
\\
To appreciate the crucial role that the thermodynamic potential plays, recall that at equilibrium the thermodynamic potential is the quantity that is minimised and from which all thermodynamic quantities are computed.  It should therefore be clear that for a full understanding of the equilibrium properties of these systems it is imperative to have as complete computation as possible of the grand canonical partition function. In particular one should recognise that the vacuum energy may also yield magnetic field dependent contributions.  \\
\\
An equivalent way of phrasing this, in the language of quantum field theory, is as follows: To find the full magnetic field dependence of the grand canonical potential, one has to compute the effective action of the electromagnetic field by integrating out the fermions in the presence of a chemical potential.  This automatically takes care of all the contributions, including vacuum, of the fermions.   When doing this, one treats the electromagnetic field as classical, i.e., one does not integrate over it.  The computation of the relevant determinants, even with classical electromagnetic field configurations, is in general intractable.  However, these determinants can be computed when restricting to stationary and spatially constant field configurations, which is the general mean field approximation made in all the works mentioned above.  This was also the approach first taken in \cite{cangemi}, but not including any anomalous couplings.  Of course, when following this approach one has to face ultraviolet divergences, which we address in some detail here.\\
\\
As far as we can establish, no computation of the effective action for the electromagnetic field in the presence of anomalous couplings has been done. The aim of this paper is to address this, particularly in the case of neutrally charged fermions that couple anomalously to the electromagnetic field.  The motivation for studying this particular case is that it is the most relevant scenario for neutron rich nuclear matter in strong magnetic fields as found in neutron stars.  Furthermore, the extension of these results to charged, anomalously coupled fermions is straightforward.  Apart from these motivations, current literature \cite{ferrer,pot,ferrer2} still contain several controversies, especially around the issue of anisotropic pressure.  We indicate a possible resolution of this controversy here, but, since this is not the main aim of this article, a more careful and detailed analysis will be carried out elsewhere.\\
\\
Once the effective action has been derived, one can extract the full magnetic field dependence of the grand canonical potential, which can then in turn be used to study the equilibrium and, at zero temperature, the ground state properties of the system.   
\section{Formalism}
A system of magnetised neutrons is described by
\begin{eqnarray}
				{\cal L}&=&
				\bar{\psi}{ (x)}
				\left[
				i\gamma^{\mu}\partial_{\mu}
				-\frac{g_b}{2}F^{\mu\nu}\sigma_{\mu\nu}-m\right]\psi{ (x)}-\frac{1}{4}F^{\mu\nu}F_{\mu\nu}	\label{calL}
\end{eqnarray}
where 
$\psi$
is the neutron 
field operator 
and $\sigma^{\mu\nu} = \frac{i}{2}\left[\gamma^\mu,\gamma^\nu\right]$ are the generators of the Lorentz group. In this paper we will use the Heaviside-Lorentz units and the convention of Ref. \cite{itzykson} for the $\gamma$-matrices (for exact expressions of quantities and their units in this conventions please see Ref. \cite{dienerPhD}). \\
\\
As already mentioned, the coupling between the dipole moment of the baryons and the electromagnetic field tensor is reminiscent of the coupling arising from the anomalous correction to the electron magnetic dipole moment. This 
is a higher order quantum correction 
due to radiative corrections to the photon-electron vertex. Since neutrons are not fundamental particles this coupling does not represent such a quantum correction, but rather an effective coupling between the neutron's magnetic dipole moment arising from its internal structure and the magnetic field. 
However, in keeping with convention we will still refer to this coupling as the ``anomalous'' magnetic moment. The coupling strength $g_b$ is therefore the neutron's magnetic dipole moment in units of fermi. \\
\\
Our aim is calculate the grand canonical partition function and grand canonical potential, or more simply the grand potential, 
of this system using the Feynman path integral approach, as described in Ref.\ \cite{kapgale}. 
The grand canonical partition function is expressed as \cite{kapgale}
\begin{eqnarray}
	{\cal Z} = \int[d\pi]\int[d\phi]\exp 
		\left[
			\int^{\beta}_{0}d\tau\int d^3 x\  i\pi \frac{\partial\phi}{\partial\tau} - {\cal H}(\phi,\pi) - \mu{\cal N}(\phi,\pi)
		\right]\label{fZ}
\end{eqnarray}
with $\beta =(k_B T)^{-1}$ the inverse temperature times Boltzmann's constant, $\tau = -it$ the analytic continuation of time in the complex plane, ${\cal H}$ the Hamiltonian density, and ${\cal N}$ the conserved density. The functional integration over $\phi$ represents all fields, while that over $\pi$ their conjugates.\\
\\
We assume that the magnetic field is a stationary and spatially constant background field pointing in the $z$-direction. Based on these assumptions we choose the gauge field to $A^\mu=(0,0,Bx,0)$. This choice simplifies  $-\frac{1}{4}F^{\mu\nu}F_{\mu\nu}$ in (\ref{calL}) to $-\frac{1}{2} B^2$ and reduces the anomalous coupling to
\begin{eqnarray}
		-\frac{g_b}{2}F^{\mu\nu}\sigma_{\mu\nu} = g_b{\bm\Sigma}\cdot {\bm B},
\end{eqnarray}
where ${\bm\Sigma} = {\bm \sigma}\otimes {\bm 1_2}$, ${\bm \sigma}$ are the Pauli matrices, and ${\bm 1_2}$ the $2\times 2$ identity matrix. Since the magnetic field is considered to be a classical background field only the fermion fields are integrated over in $\cal Z$. \\
\\
The Hamiltonian density is
\begin{eqnarray}
				{\cal H}=\psi^{\dagger}\left(i\frac{\partial}{\partial t}\right)\psi-{\cal L}=
				\bar{\psi}
				\left[
				-i{\bm\gamma}\cdot{\bm\nabla}
				-g_b{\bm \Sigma}\cdot {\bm B}
				+m\right]\psi
				+\frac{1}{2}{\bm B}^2,\label{hamil}
\end{eqnarray}
and the conserved density in (\ref{fZ}) is the neutron number density. Thereby $\cal Z$ reduces to
\begin{eqnarray}
	{\cal Z} = \int[id\psi^{\dagger}][d\psi]
	\exp 
		\int^{\beta}_{0}d\tau
			\int\! d^3 x\,\bar{\psi}
				\left( 
					\gamma^0\frac{\partial}{\partial\tau} 
					+ i{\bm\gamma}\cdot{\bm\nabla} 
					+ g_b{\bm \Sigma}\cdot {\bm B} 
					- m  
					- \mu \gamma^0
				\right)\psi
				-\frac{{\bm B}^2}{2}. 
\end{eqnarray}
From here on we will momentarily ignore the contribution of the Maxwell term since this only adds a constant factor. Since the addition of the anomalous term does not affect the commutation relations of the baryon fields, the method of Ref.\ \cite{kapgale} can be used to evaluate the functional integrals. Thereby ${\cal Z}$ reduces to the determinant of the matrix $D$ which does not contain any field operators,
\begin{eqnarray}
	D=-i\beta
			\left[
				\left(
					-i\omega_n+\mu
				\right)-
				\left(
					i\gamma^0{\bm\gamma}\cdot{\bm\nabla} 
					- g_b{\bm \Sigma}\cdot {\bm B} 
					+ m  
				\right)
			\right],
\end{eqnarray}
so that \cite{kapgale} 
\begin{eqnarray}
	\mbox{ln } {\cal Z} = \mbox{ln Det } D = \mbox{Tr ln } D
	= \sum_\lambda \sum_n \sum_{\bm k}\mbox{ln } 
		\left\{ 
			\beta^2
			\left[
				\left(
					\omega_n + i\mu
				\right)^2
				+\omega^2
			\right]
		\right\},\label{sumZ}
\end{eqnarray}
where $\omega_n =(2n+1)\,T$ are the Matsubara frequencies. The $\omega$ is the spectrum of the Hamiltonian (\ref{hamil}) which is known from Ref.\ \cite{dienerPhD} (and references therein):
\begin{eqnarray}
	\omega({\bm k},\lambda)
	&=& \pm\sqrt{\left(\sqrt{k_{\bot}^2+{m}^2}+\lambda g_b B\right)^2+k_{z}^2} \label{singlepatE},
\end{eqnarray}
where
\begin{itemize}
	\item $\lambda = \pm 1$ distinguishes the different orientations of the neutron dipole moment, and
	\item 
	$k_{\bot}^2 = k_{x}^2+k_{y}^2$, perpendicular to ${\bm B} = B{\hat z}$, given the choice of $A^\mu$.
\end{itemize}
The sum over $n$ is computed in \cite{kapgale} and thereafter (\ref{sumZ}) reduces to
\begin{eqnarray}
	\mbox{ln}\, {\cal Z} 
	= V\sum_\lambda \int\frac{d^3 k}{(2\pi)^3}
			\left[
				\beta \omega+
				\mbox{ln}
				\left(
					1+e^{-\beta(\omega-\mu)}
				\right)+
				\mbox{ln}
				\left(
					1+e^{-\beta(\omega+\mu)}
				\right)
			\right]-\beta V\frac{B^2}{2}\label{Z}
\end{eqnarray}
where we have again included the Maxwell contribution of the magnetic field. The grand potential is defined as 
\begin{eqnarray}
	\Omega(\mu, T) = -\frac{\mbox{ln}\, {\cal Z}}{V \beta},
\end{eqnarray}
and its various thermodynamic relations are \cite{kapgale} 
\begin{eqnarray}
	\frac{S}{V} = -\left(\frac{\partial\Omega}{\partial T}\right)_{\mu},
	\ \rho = \frac{N}{V} = -\left(\frac{\partial\Omega}{\partial \mu}\right)_{T}.
\end{eqnarray}
Due to the presence of the magnetic field the spherical symmetry of the ground state is broken and the pressure is generally anisotropic. The longitudinal $P_{\parallel}$ and transverse $P_{\perp}$ components of the pressure are given by \cite{ferrer}
\begin{eqnarray}
	P_{\parallel} = -\Omega,\  P_{\perp} = -\Omega+B\frac{d\Omega}{dB}.\label{presb}
\end{eqnarray}
In the limit of $\beta\rightarrow\infty$, so that $T\rightarrow 0$, for $\mu\neq0$ (and $\mu>m$) $\Omega$ becomes 
\cite{cangemi}
\begin{eqnarray}
	\Omega
	= \sum_\lambda \int\frac{d^3 k}{(2\pi)^3}
				\big\{
					-\omega+
					\left(
						\omega-\mu
					\right)
					\,\Theta\!
					\left[
						\,\omega-\mu
					\right]
				\big\}
				+\frac{B}{2}.\label{delO}
\end{eqnarray}
Using the following relation between the various thermodynamic quantities, $dE = TdS-PdV+\mu N$, the energy density of the system is 
\begin{eqnarray}
	\epsilon = \frac{E}{V} = \sum_\lambda \int\frac{d^3 k}{(2\pi)^3}
				\big(
					-\omega	
					+\omega
					\,\Theta\!
					\left[
						\,\mu-\omega
					\right]		
				\big)
				+\frac{B}{2}.\label{eps}
\end{eqnarray}
Therefore the first term of the integrand in (\ref{delO}) corresponds to the zero-point energy of the vacuum, while the second term is the contribution from all the filled positive energy particle states (since $T = 0$ the chemical potential $\mu$ is equal to the neutron's Fermi energy).\\
\\
Since we are interested in the dependence of $\Omega$ on the magnetic field $B$, all terms that are independent of $B$ are irrelevant. In particular this implies that the only relevant quantity is the difference between the grand potential for $B\neq 0$ (magnetised) and $B=0$ (unmagnetised), which we expand in a power series in $B$. The difference in the contributions from the occupied positive energy neutron states is finite and calculable for a fixed number of neutrons and we will refer to it as $\Delta(\mu,B)$. The difference in $\Omega$ for magnetised and unmagnetised neutrons is therefore written as 
\begin{eqnarray}
	\Omega(B)-\Omega(0) = \Delta(\mu,B) + \sum^{\infty}_{n=0}a_{2n}(\Lambda)B^{2n},\label{epsdelta}
\end{eqnarray} 
where $\Lambda$ is some (large) momentum cut-off that in the end will be taken to infinity. The expansion coefficients $a_{2n}$ are determined by expanding the single particle energy (\ref{singlepatE}) for $B\neq0$  
\begin{eqnarray}
	\sum_\lambda\omega(B)
	= 
	\sum_\lambda\omega(0) 
	+ \frac{g_b^2k_z^2B^2}{4\left(k^2+m^2\right)^{3/2}}
	-\frac{g_b^4k_z^4\left(5k_z^2-4\left(k^2+m^2\right)\right)B^4}{8\left(k^2+m^2\right)^{7/2}}
	+{\cal O}(B^6).\label{expe}
\end{eqnarray}
The first term in the expansion is just the zero-point energy 
for unmagnetized neutrons and therefore will be cancelled in (\ref{epsdelta}). In the remaining terms the $k_\perp$ integrals can be performed without any divergences appearing, while a momentum cut-off $\Lambda$ is introduced for the $k_z$ integrals. Dropping terms of order $\Lambda^{-1}$ and higher yields,  
\begin{eqnarray}
	&&\sum^{\infty}_{n=0}a_{2n}(\Lambda)B^{2n} = \int\frac{d^3 k}{(2\pi)^3}\left(\sum_\lambda\omega(B)-\sum_\lambda\omega(0)\right)\label{O4}\\
	&&=\left(\frac{g_b^{2}\Lambda^2}{4\pi^2}+\frac{g_b^2m^2\left(1-\log[4]\right)}{8\pi^2}+\frac{g_b^2m^2}{4\pi^2}\log\left[\frac{m}{\Lambda}\right]\right)B^2
	+\left(\frac{g_b^4\log[2]}{24\pi^2}+\frac{g_b^4}{24\pi^2}\log\left[\frac{m}{\Lambda}\right]\right)B^4\nonumber,
\end{eqnarray}
Furthermore $B<1\ (\mbox{fm}^{-2})$ and therefore higher order terms in $B$ become irrelevant. 
We note the appearance of quadratic and logarithmic divergences when  $\Lambda \rightarrow\infty$.  As we do not have any physical renormalisation point, we remove the infinities through  a minimal subtraction scheme \cite{brown}.  In addition, we must keep in mind that as $\Omega$ here represents the full grand potential of matter and fields, it must yield, upon setting $g_b=0$, the contribution of the Maxwell term $\frac{1}{2}B^2$ and free fermions, i.e., $\Delta(\mu,B) = 0$, which implies that $a_2 = \frac{1}{2}$ and $a_{2n} = 0,\  \forall n\neq1$ at this point.  Then the magnetic field dependent part of the grand potential for magnetised neutrons is 
\begin{eqnarray}
	\Omega &=& 
	\sum_\lambda \int\frac{d^3 k}{(2\pi)^3}
			(-\beta^{-1})\left[
				\mbox{ln}
				\left(
					1+e^{-\beta(\omega-\mu)}
				\right)+
				\mbox{ln}
				\left(
					1+e^{-\beta(\omega+\mu)}
				\right)
			\right]\nonumber\\
		&\ &+\frac{B^2}{2}\left(1-\frac{g_b^2m^2\left(1-\log[4]\right)}{4\pi^2}\right)
		-\frac{g_b^4B^4\log[2]}{24\pi^2}.\label{OBbeta}
\end{eqnarray}
In the zero temperature limit of $\beta\rightarrow\infty$ (\ref{OBbeta}) becomes
\begin{eqnarray}
	\Omega &=& 
	\sum_\lambda
		\int\frac{d^3 k}{(2\pi)^3}
		\ \omega\, \Theta\!
		\left[
			\,\mu-\omega
		\right]
		-\mu\rho
		+		
		\frac{B^2}{2}
		\left(
			1-\frac{g_b^2m^2\left(1-\log[4]\right)}{4\pi^2}-\frac{g_b^4B^2\log[2]}{12\pi^2}\right),\label{OB}
\end{eqnarray}
where $\rho = \frac{N}{V}$ and
\begin{eqnarray}
	N &=& 
	V\sum_\lambda
		\int\frac{d^3 k}{(2\pi)^3}
		\Theta\!
		\left[
			\,\mu-\omega
		\right].
\end{eqnarray}
Not surprisingly we note that the fermion contribution to the effective action for the electromagnetic field modifies the coefficient in front of the $B^2$ term, which simply implies that the vacuum is polarised due to the presence of the matter fields.   
\section{Discussion}\label{sec:dis}
\begin{figure}[tbt]
	\centering
		\includegraphics[width=.750\textwidth]{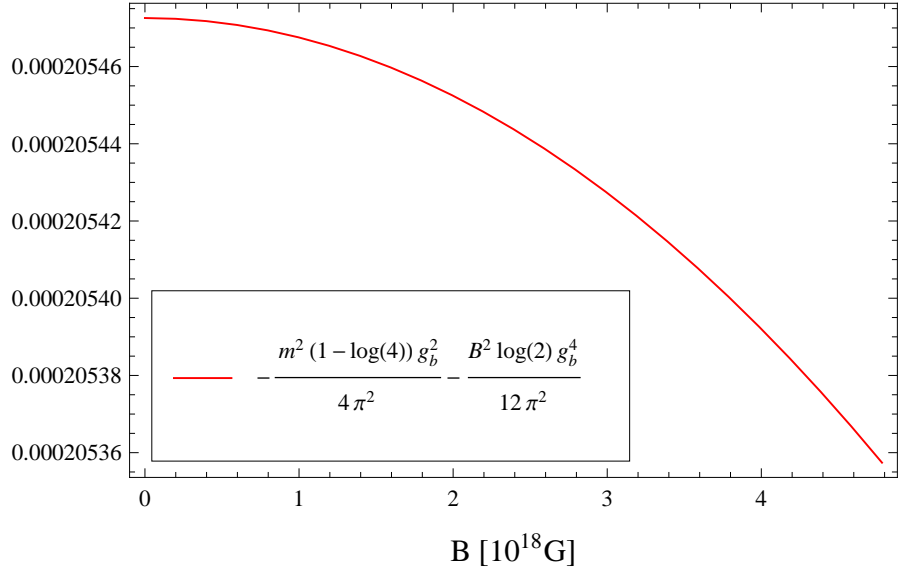}
	\caption{ (Color online) The contribution of the zero-point energy to the last term in (\ref{OB}). Correspondingly it should be compared to 1, which represents the contributions from the magnetic field. 
	}
	\label{fig:vac}
\end{figure}

The zero-point energy is the infinite contribution from the filled negative energy (antiparticle) states that form the vacuum. For most dense matter systems this ``sea'' of antiparticles is ignored since only the relative changes in the filled positive energy states are assumed to have physical consequences. Generally this is a good approximation for dense nuclear matter system since the high value of the Fermi energy, $\omega_F\gg-m$, blocks any low energy vacuum excitations due to the filled positive energy states. Figure \ref{fig:vac} demonstrates the effect of the vacuum contribution, using the free value of $g_b = \frac{1}{2}\, 1.913\mu_N = g_b^{(0)} \approx 0.0305$ fm, to the thermodynamic potential and confirms its relative smallness.
\\
\\
It should, however, be kept in mind that since we treated the electromagnetic field as classical, photon loop corrections to $g_b$, and therefore also medium effects, have not been accounted for.  These may significantly alter the value of $g_b$ and therefore also the effect of the vacuum contribution.  Indeed, there is good experimental evidence that the neutron dipole moment may increase with density: in Copper isotopes the dipole moment changed by about 50\% over an atomic mass range of 10 \cite{stone}.  This suggests that the contribution of the vacuum energy may be much more relevant if all radiative corrections, and thus medium effects, are more carefully accounted for, which is one of the main motivations for computing these corrections here.   One can, of course, simply treat $g_b$ as a parameter and investigate its potential impact on the equilibrium properties of these systems.  This suggests considerable quantitative and qualitative effects and may even, if $g_b$ is made large enough, lead to a ferromagnetic phase \cite{dienerPhD}.  These are, however, currently speculations that need to be scrutinised carefully in future.\\
\\
A final observation we make here relates to the rather controversial issue of the anisotropic pressure.  Recall that at equilibrium $-\Omega=-\Omega_{\rm f}+\frac{B^2}{2}$, with $\Omega_{\rm f}$ the fermion contribution, is minimised or, alternatively, the entropy is maximised at fixed temperature and density.  From this one can compute the equilibrium value of the magnetic field as 
\begin{equation}
B=\frac{\partial \Omega_{\rm f}}{\partial B}=M,
\end{equation}
where $M$ is the magnetisation of the fermion gas.  Substituting this in (\ref{presb}) shows that at equilibrium
\begin{equation}
P_{\parallel} = P_{\perp} =-\Omega_{\rm f}+\frac{M^2}{2}.
\end{equation}
This suggests that an anisotropic pressure may be a signal that the system is out of equilibrium and, since $B\neq M$, the presence of free currents and a dynamo effect.  These are merely hints that warrant further investigation elsewhere.
\section{Conclusion}
In this article we have computed the full magnetic field dependence of the grand potential for magnetised neutron matter coupled anomalously to the magnetic field.  This includes vacuum effects but not yet radiative photon corrections.    We did this by computing the effective action of the electromagnetic field, treated as a classical field, by integrating out the fermions.  The result can be used to compute the properties of these systems at equilibrium and this may shed light on the controversial issues of anisotropic pressure and the possibility of a ferromagnetic phase transition. 
\section{Acknowledgements}
This research is supported by the National Research Foundation of South Africa.

\bibliography{bibeffecA}

\end{document}